\documentclass[icps3]{svjour}
\usepackage {amsmath}
\usepackage{graphics}
\usepackage {epsfig}
\usepackage {longtable}
\usepackage {fancyheadings}
\usepackage{flafter}
\begin{document}
\title{Measurement of the energy dependence of phase relaxation by single electron tunneling}
\author{P.~K\"onig\inst{1}, J.~K\"onemann\inst{1}, T.~Schmidt\inst{1}, E.~McCann\inst{2}, V.~I.~Fal'ko\inst{2} \and R.~J.~Haug\inst{1}}
\institute{Universit\"at Hannover, Inst.~f.~Festk\"orperphysik, Appelstr. 2, 30167 Hannover, Germany
\and School of Physics \& Chemistry, Lancaster University, LA1 4YB Lancaster, United Kingdom}  
\sloppy
\maketitle
\renewcommand{\abstractname}{Abstract}
\begin{abstract}
Single electron tunneling through a single impurity level is used to probe the fluctuations of the local density of states in the emitter. The energy dependence of quasi-particle relaxation in the emitter can be extracted from the damping of the fluctuations of the local density of states (\emph{LDOS}). At larger magnetic fields Zeeman splitting is observed. 
\end{abstract}
  \section{Introduction}
\begin{figure}[!h]
\centerline{\epsfig{file=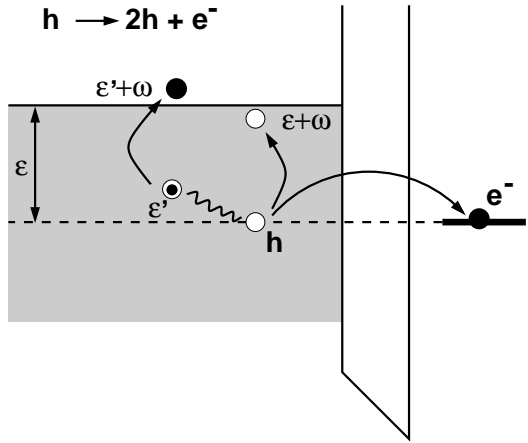,width=7cm}}
\caption{Sketch of quasi-hole relaxation.}
\label{relax}
\end{figure}

Relaxation is an essential ingredient for understanding quantum transport  phenomena in solid state devices. It can be described as the decay of elementary excitations in the quasi-particle picture of Fermi liquids \cite{landau}. 
A typical process is  a decay of one quasihole in one quasi-electron and two quasi-holes, sketched in Fig. \ref{relax}. 
The relaxation rate $\gamma$ of such elementary excitations in clean degenerate Fermi liquids  is given by  phase space arguments as $\gamma\propto\epsilon^2$, with the excitation energy $\epsilon$ measured relative to the Fermi level. 
In disordered systems the theory for Fermi liquids predicts a different power law $\gamma\propto\epsilon^{3/2}$ \cite{altshuler,sivan_theory}. 
In a previous experiment a small quantum dot was used as a sensitive spectrometer of such relaxatiom processes in a large disordered quantum dot \cite{sivan_exp}.

\section{Experiment}
The experiment was performed with a strongly asymmetric double-barrier heterostructure, grown on  $n^+$-type GaAs. The structure consists of a 10 nm wide undoped GaAs quantum well and two Al$_{0.3}$Ga$_{0.7}$As tunneling barriers of 5 nm and 8 nm width.  Two 7 nm wide GaAs-spacer layers separate the nominally undoped  active region  from the 300 nm thick highly doped GaAs contact layers. The charge carrier concentration at the barrier was found to be 3.3$\times$10$^{17}$cm$^{-3}$. From this heterostructure we fabricated a pillar with 2 $\mu$m diameter,  which contains some impurities inside the quantum well. 
\begin{figure}[!h]
\centerline{\epsfig{file=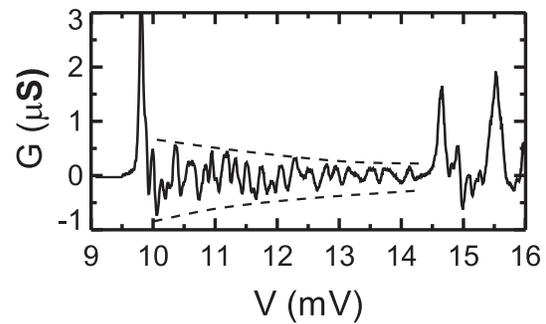,width=7cm}}
\caption{Typical differential conductance versus applied voltage (T=20 mK)}
\label{gkenn}
\end{figure}
Fig. \ref{gkenn} shows the differential-conductance versus voltage characteristics in the low field regime (at B=0.1 T). One clearly distinguishes three large main peaks of the differential conductance at V=9.8 mV, V=14.6 mV and V=15.6 mV, which can be attributed to single electron tunneling (\emph{SET}) through the lowest impurity states inside the quantum well of the sample.\\
The electrons tunnel first through the thick tunnel barrier into the energetically lowest discrete states in the quantum well, so the tunneling current is limited by the thick, less transparent tunnel barrier. As the tunneling current itself is proportional to the density of states $I\propto\nu$, the current-voltage characteristics displays the energy-dependent local density of states (\emph{LDOS}) of the occupied emitter states with respect to the emitter Fermi Level \cite{euro}. Therefore the differential conductance $G=dI/dV$ images the derivation of the LDOS with respect to energy $G\propto d\nu/dE$.\\ 
 Between the two first large peaks one clearly resolves an oscillatory fine structure, which is found to be temperature independent. This fine structure displays the mesoscopic fluctuations of the \emph{LDOS}. To perform a statistical analysis, we measured current-voltage characteristics between [0-15 T] with a stepwidth of 10 mT. \\
\begin{figure}[!h]
\centerline{\epsfig{file=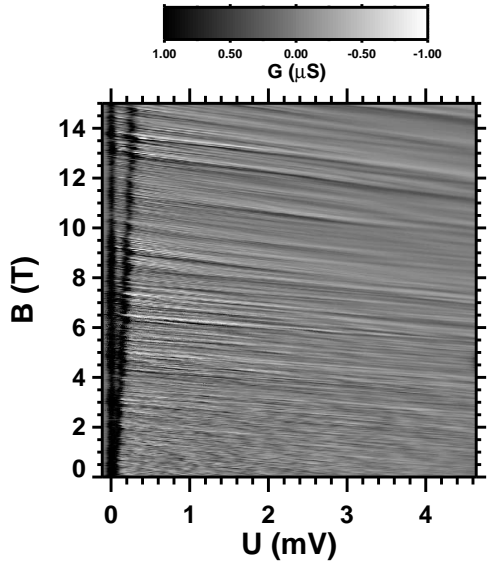,width=7cm}}
\caption{Differential conductance G(V,B), aligned to the field-dependent main peak position in the magnetic field range of [0-15 T] shown in grey scale}
\label{field}
\end{figure}
 Fig. \ref{field} shows the voltage and magnetic field dependence of the differential conductance at 20 mK starting from the onset of the current (first large peak in the differential conductance V=9.8 mV  at B=0 T). The G-V-curves are all aligned to the first conductance peak, which is set to 0 meV corresponding to the Fermi level. \\ 
In the low field regime up to 2.5 T ($\omega_C\tau<1$) an irregular pattern is seen. At approximately 2.8 T one observes the formation of stripes \cite{high}, which become more clearly pronounced at higher B-fields. 

\section{Discussion}
As a measurable quantity for the LDOS fluctuations we investigate the variation of the differential conductance var$_BG$ = $<\delta G(B)^2>_B$, with $\delta G$ = $G(B)$ - $<G>_B$ in the field range [0-1 T], where the Landau quantization can be neglected with respect to disorder, i.e. in the classical limit $\omega_C\tau$ $<$ 1 with $\tau$=0.14 ps. In this limit, the LDOS fluctuations are related to the dephasing rate $\gamma$ as \cite{falko}:
\begin{equation}
\mbox{var}_BG= G_N^2/(1+\hbar\gamma/\Gamma)^{3/2}\label{fra}.
\end{equation}
Here, the spectrometer width $\Gamma$ is given through the width of the main conductance peak at $V_S$=9.8 mV as 36 $\mu$eV \cite{phase}. The pre-factor $G_N$ could be extracted from the condition, that the quasi-particle relaxation rate $\gamma$ (proportional to the phase space volume of the final scattering states) approaches 0, when the excitation energy $\epsilon$ goes to 0, e.g. $V\rightarrow V_S$.\\
\begin{figure}[!h]
\centerline{\epsfig{file=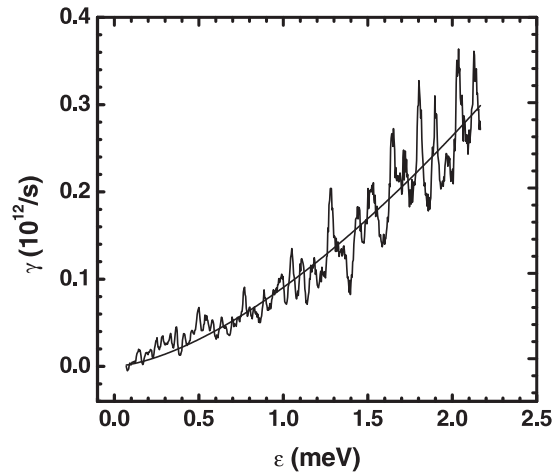,width=6.2cm,angle=270}}
\caption{Experimentally extracted quasi-particle relaxation rate $\gamma$ with a fit $\gamma=a\cdot\epsilon^b$, yielding a=0.091 $\cdot 10^{12}$ $s^{-1}$/meV$^b$, b=1.54.}
\label{rate}
\end{figure}
With the help of eq. \ref{fra} we extracted  the energy dependence of $\gamma$, shown in Fig. \ref{rate}. A two-parameter power law fit of the data yields an exponent of 1.54 in very good agreement with the theory of highly disordered conductors. In such systems like the investigated highly-doped semiconductor, the diffusive motion of electrons favors small momentum transfer, which leads to a quasi-particle relaxation rate $\gamma\propto\epsilon^{3/2}$ \cite{altshuler}.\\
At intermediate magnetic fields the oscillatory pattern in Fig. \ref{field} starts to disappear and long stripes of LDOS resonances evolve. They result from Landau quantization which reduces the lateral quantum diffusion in the highly disordered emitter. The disordered states in the emitter assume now a quasi-onedimensional character, so that a fan-like structure of the LDOS resonances appears. At higher magnetic fields these resonances appear to be more pronounced at small voltages, whereas they seem to wash out at high bias voltages. One determines the slope of the resonances beyond the magnetic quantum limit at B=11.4 T, where only one spin degenerate Landau band is populated, to be $dV/dB=3.0$ mV/T, which is roughly consistent with the increase of the cyclotron energy for GaAs (1.73 meV/T) with magnetic fields (after energy conversion with $\alpha$=0.5).\\
The splitting of the main differential conductance peak at 0 meV in Fig. \ref{field} is attributed to the Zeeman splitting of the impurity state \cite{deshpande}. A linear fit yields an effective $g$-factor of the impurity S $|g_S^*|=0.14$ \cite{spin}.\\
The main features of the spin-up and spin-down components of the \emph{LDOS} are found to be identical apart from Zeeman splitting \cite{high}. Such comparisons for fixed energies and varying magnetic fields are possible due to the quasi-one-dimensional nature of the density of states in high magnetic fields and due to the depopulation of the Landau-levels  


\end{document}